\documentclass[12pt]{iopart}

\usepackage{amssymb,color}

\newcommand{\be}{\begin{equation}}
\newcommand{\en}{\end{equation}}

\newcommand{\pap}{Ref.\cite{YO} } 
\newcommand{\acoef}{A_{\alpha_1\alpha_2\alpha_3}}

\newcommand{\tb}{\tilde{\beta}}

\newcommand{\xx}{x^{2(\epsilon-1)}}
\newcommand{\yy}{y^{2(\epsilon-1)}}
\newcommand{\xy}{(x+y)^{2(\epsilon-1)}}
\newcommand{\re}{{\rm Re}\,}
\newcommand{\dptau}{\frac{\partial \Pi_2}{\partial \ln\tau}}
\newcommand{\prl}{Phys. Rev. Lett.}
\newcommand{\pre}{Phys. Rev. E}
\newcommand{\prb}{Phys. Rev. B}

\begin{document}
\title{Return probability and scaling exponents in the critical random matrix ensemble}
\author{V.E. Kravtsov$^1$, A. Ossipov$^2$, O.M. Yevtushenko$^3$
}
\address{$^1$Abdus Salam ICTP, P.O. 586, 34100 Trieste, Italy \\
                 $^2$School of Mathematical Sciences, University of Nottingham, Nottingham
                          NG7 2RD, United Kingdom \\ 
                 $^3$Arnold Sommerfeld Center and Center for Nano-Science,
                          Ludwig-Maximilians-University, Munich D-80333, Germany\\
}

\begin{abstract}
We study an asymptotic behavior of the return probability for the critical random 
matrix ensemble in the regime of strong multifractality. The return probability
is expected to show critical scaling in the limit of large time or large system size.
Using the supersymmetric virial expansion we confirm the scaling law and find 
analytical expressions for the fractal dimension of the wave functions $d_2$ and 
the dynamical scaling exponent $\mu$. By comparing them we verify the validity 
of the Chalker's ansatz for dynamical scaling.
%
%
\end{abstract}

\maketitle

\section{Introduction}

It is well-known that 
the wave functions at the point of the Anderson metal-insulator transition are {\it fractal} \cite{Wegner}. 
Their amplitudes exhibit self-similar fluctuations at different spatial scales. The standard way to quantify such a complicated behavior is to consider the scaling of moments of the wave functions $\psi_n({\bf r})$  with the system size $L$:
\be
\label{IPR} I_q=\sum_{\bf r}\left\langle
|\psi_n({\bf r})|^{2q}\right\rangle\propto L^{-d_q(q-1)},
\en
where $\langle\dots\rangle$ stands for averaging over disorder realizations 
 and  over a small energy window.
The fractal dimension $d_q$, which is different from zero and from the dimensionality of the space $d$, is a fingerprint of the fractal wave functions. For the {\it multifractal} wave functions $d_q$ depends non-trivially on $q$, thus an infinite set of scaling exponents is required for the full description of the wave functions in this case.

Additionally to the non-trivial scaling of the moments of the critical wave functions taken at a fixed energy, the correlations of wave functions at different energies show the critical behavior as well. The simplest correlation function involving two eigenstates corresponding to two different energies  $E_m $ and $ E_n $ can be defined as 
\be\label{mat-elem} 
C(\omega)=\sum_{{\bf r}} 
   \langle |\psi_{n}({\bf
      r})|^{2}\,|\psi_{m}({\bf r})|^{2}\,\delta(E_{m}-E_{n}-\omega)
   \rangle. 
\en 
As any other correlator at criticality $C(\omega)$ is expected to decay in a power-law fashion
\be \label{C(omega)}
  C(\omega)\propto
  (E_{0}/\omega)^{\mu},\quad  \Delta \ll \omega \ll  E_{0},
\en 
where $\Delta$ is the mean level spacing and $E_0$ is a high-energy cutoff. What is more surprising is the fact that the dynamical exponent $\mu$ is related to the fractal dimension $d_2$ in a simple way
\be
\label{Chalk-anz}
\mu=1-d_2/d.
\en
This relation  was suggested by Chalker and Daniel \cite{ChD88, Chalk} and confirmed by a great number  of 
computer simulations \cite{ChD88,CK07,HSch94} thereafter. As $ \, E_0/\omega \gg 1 \, $ and $ \, \mu > 0 $,
Eq.(\ref{C(omega)}) implies an enhancement of correlations in critical systems \cite{CK07} which is possible 
only if there is a strong overlap of different wave functions. This rather counterintuitive picture becomes particularly 
striking in the regime of strong multifractality $d_2\ll d$, where eigenstates are very sparse \cite{FM-Crit} and 
almost localized.  We are not aware of any analytical calculations supporting its validity in this case.

The critical enhancement of the correlations plays an important role in the theory of interacting systems
(cf. ``multi-fractal'' superconductivity \cite{SupCond} and the Kondo effect \cite{Kondo}) and, therefore, 
the theory of the critical correlations still attracts a considerable attention in spite of its long history.

%
%

The aim of the present work is to demonstrate that the critical scaling holds true and confirm 
the validity of the dynamical scaling relation (\ref{Chalk-anz}) with accuracy up to the second order of the 
perturbation theory. Some results of this work have been announced in a brief form in Ref.\cite{KOYC10}.

%
%


The knowledge of the second order perturbative results 
is extremely important because it allows one to confirm the critical scaling. 
Besides, subleading terms in the scaling exponents can reveal some model-dependent features in contrast
 to the leading order perturbative result which is universal for a wide class of different critical models~\cite{FOR09,ROF11}.

%
%

In this paper we consider a particular model relevant for the Anderson metal-insulator transition -- the power law random banded matrix ensemble \cite{MFD96}. The matrix elements $H_{mn}$ of the Hamiltonian are given by the independent, Gaussian distributed random variables with the only constraint that matrix $H$ is Hermitian. Their mean values are equal to zero and the variances are defined as
\begin{eqnarray}\label{Ham}
 \langle |H_{mn}|^2\rangle =\frac{1}{2}\left\{\begin{array}{ll} 1, \quad \quad \quad n=m \\ \frac{b^2}{(n-m)^2}, \;\;|n-m|\gg {\rm max} \{ b, \, 1 \} \end{array}\right.
\end{eqnarray}
The long-range power-law decay of typical matrix elements  $|H_{mn}|^2$ leads to the critical behavior at any value of the parameter $b$. This allows one to study the model perturbatively either for $b\gg1$ or for $b\ll1$. The latter condition corresponds to the regime of strong multifractality investigated in the present work. In this regime the both scaling exponents $\mu$ and $d_2$ can be expanded into the power series in $b$ and then compared term by term. 

The structure of the paper is as follows. In Section \ref{sec-ret-prob} we give an equivalent formulation of the Chalker's ansatz in terms of the return probability. In Section \ref{sec-P2} the integral representation for the first order result for the return probability is derived. It is used in Section \ref{sec-exp-first} to calculate the scaling exponents in the first order in $b$. The second order result for the return probability is discussed in Section \ref{sec-P3}. The corresponding second order expressions for the dynamical exponent $\mu$ and fractal dimension $d_2$ are derived in Section \ref{sec-mu} and Section \ref{sec-d2} respectively.
%
%
All calculations in Sections \ref{sec-P2}--\ref{sec-d2} are done for the unitary symmetry class. 
A brief announcement of analogous results for the orthogonal case is presented in  \ref{GOE-2Q}; more 
detailed study will be published elsewhere.

\section{Return probability and scaling exponents}\label{sec-ret-prob}

It is convenient to reformulate the Chalker's ansatz (\ref{Chalk-anz}) in terms of the return probability
\be \label{RP}
P_N(t)=\int_{-\infty}^{\infty} d\omega\: e^{-i\omega t} C(\omega),
\en
where a matrix size $N$ plays the role of the system size. Using the definition of $C(\omega)$ (\ref{mat-elem})
it is easy to show that in the limit $t\to\infty$ the return probability tends to a finite limit, 
which is nothing else than the inverse participation ratio $I_2$
\be\label{P(L)}
    \lim_{t/N\to \infty} P_{N}(t)=I_2\propto N^{-d_{2}}.
\en
Thus the knowledge of the return probability gives a way to calculate the fractal dimension $d_2$. On the other hand considering the limit $N\to\infty$ at a fixed large time $t$ one finds that
\be\label{P(bt)}
\lim_{N/t\to \infty}P_{N}(t)\propto t^{\mu-1},
\en
as it follows from Eq.(\ref{C(omega)}). So the dynamical scaling exponent $\mu$ can be also extracted from the behavior of the return probability. Thus we conclude that the Chalker's ansatz (\ref{Chalk-anz}) is equivalent to the following statement
\be\label{d-log} 
\mu-1=\lim_{t\to\infty}\frac{\partial}{\partial\ln t}\lim_{N/t\to\infty}\ln P_N(t) =
\lim_{N\to\infty}\frac{\partial}{\partial\ln N}\lim_{t/N\to\infty}\ln P_N(t)=-d_2.
\en
In the regime of strong multifractality the return probability can be calculated perturbatively using the method of 
the virial expansion in the number of resonant states, each of them being localized at a certain site n. The virial 
expansion formalism was developed in Ref.\cite{YK03} following the initial idea of Ref.\cite{Levitov}.
The supersymmetric version of the virial expansion \cite{YO,KYC} is formulated in terms of integrals over
super-matrices. In particular, it allows us to represent $ \, P_N(t) \, $ as an infinite series of integrals over an 
increasing number of super-matrices associated with different sites:
%
%
\be 
P_N(t)=P^{(1)}+P^{(2)}+P^{(3)}+\dots,
\en
with $P^{(1)}=1$ and $P^{(i)}=b^{i-1}f^{(i)}(bt)$.
Functions $f^{(j)}$ are governed by a hybridization of $ \, j \, $ localized states and can be calculated explicitly 
by means of integrals over $ \, j \, $ different supermatrices.
%
%
The above expansion implies the corresponding expansion for $\ln P_N(t)$:
\be
\ln P_N(t)=P^{(2)}+\left(P^{(3)}-\frac{1}{2}\left(P^{(2)}\right)^2\right)+\dots,
\en 
where the first term is of the first order in $b$, two terms in the brackets are of the second order in $b$ and so on. This representation allows one to find the corresponding power law expansions for the fractal dimensions $d_2$ and the dynamical exponent $\mu$ and hence to check the Chalker's ansatz in the form of Eq.(\ref{d-log}) order-by-order.

We emphasize that the scaling exponents are finite constants and hence according to Eq.(\ref{d-log}) 
the leading at  $N \to \infty$ or $t\to\infty$ terms in $ \, \ln P_N(t) \, $ must diverge 
{\it logarithmically} with only linear in $\ln N$ or $\ln t$ terms present. Higher order terms  
of the form $\ln^m N$ or $\ln^m t \, $ $ \, (m>1)$ would generate divergent contributions to $ \, d_2 \, $ and $ \, \mu \, $ 
indicating a violation of the power law scaling. As a matter of fact, both $ \, P^{(3)} \, $ and 
$ \, \left(P^{(2)}\right)^2 \, $ do contain higher order terms, such as $\ln^2 N$ or $\ln^2 t $.
 We prove below that these divergent terms cancel out in the combination 
$ \, \left(P^{(3)}-\frac{1}{2}\left(P^{(2)}\right)^2\right) $. This is a necessary condition for the
existence of the critical scaling and of the Chalker's ansatz.
 
\section{Integral representation for $P^{(2)}$} \label{sec-P2}

The return probability $P(t)$ can be expressed in terms of the Green's functions as
\be\label{ret_prob_def}
P_N(t)=\frac{\Delta^2}{2\pi^2N}\sum_{p=1}^{N}\langle \langle {\cal G}_{pp}(t) \rangle \rangle, 
\en
where $\langle\langle ab\rangle\rangle\equiv \langle ab \rangle -\langle a \rangle \langle b \rangle$ 
and the diagonal matrix elements of time dependent correlator $ {\cal G}_{pp}(t)$ are related to its 
energy dependent counterparts  ${\cal G}_{pp}(\omega)$ by the Fourier transform
\be
   {\cal G}_{pp}(t) = \frac{1}{\Delta} \int d \omega \, e^{- \imath \, \omega t}
           \re {\cal G}_{pp}(\omega).
\en 
For the latter quantity, defined by the product of the matrix elements of the retarded and
 the advance Green's functions ${\cal G}_{pp}(\omega)\equiv G^R_{pp}(E+\omega/2)G^A_{pp}(E-\omega/2)$, 
the perturbation theory has been developed in \pap. The leading order term of the perturbation 
theory, corresponding to the diagonal part of a random matrix, is
 $\re {\cal G}_{pp}(\omega)=(2\pi^2/\Delta)\delta(\omega)$. Substituting this expression into Eq.(\ref{ret_prob_def}) yields $P^{(1)}=1$ reproducing the correct normalization of the return probability. The next order approximation taking into account an ``interaction'' between pairs of resonant states is given by Eq.(55) of \pap. The corresponding results for the return probability reads \footnote{The right hand side of Eq.(55) should be multiplied by $\sqrt{2}$, as  in the present calculations we fix $E=0$, while the averaging over $E$ was performed in \pap}
\be\label{dbl_sum}
P^{(2)}=\frac{2\sqrt{\pi}}{Nt}\sum_{n\neq p}^N\sum_{k=1}^{\infty}\frac{(-2b_{pn}t^2)^k}{(k-1)!}\frac{k}{2k-1}.
\en
This expression was derived for an arbitrary variances of the off-diagonal matrix elements $b_{pn}=\frac{1}{2}\langle |H_{pn}|^2 \rangle$; for PLBRM model $b_{pn}=\frac{1}{4}(1+|p-n|/b)^{-2}$. In the large $N$ limit, the double sum in Eq.(\ref{dbl_sum}) may be replaced by the integral:
\be\label{sum-int2}
\sum_{p=1}^N\sum_{n \ne p}^N f(|p-n|)\approx 
2\int_0^N dy\int_0^y dx f(x)\approx 2 N \int_0^N dx f(x),
\en
where the last equality is justified in  \ref{app1}. In the continuum limit, the counterpart of $b_{pn}$ is given by  $b^2/4x^2$ (which is valid for $|p-n|\gg b$), however this expression leads to the appearance of divergent integrals at $x\to 0$ and hence should be regularized. To this end we replace $b_{pn}$ by $b^2/4x^{2(1-\epsilon)}$ with $\epsilon>0$ and take the limit $\epsilon\to 0$ at the end of the calculations. Thus in the continuum limit we obtain
\be\label{p_two}
P^{(2)}=\frac{4\sqrt{\pi}}{t}\int_0^N dx \sum_{k=1}^{\infty}\left(\frac{-b^2t^2}{2x^{2(1-\epsilon)}}\right)^k \frac{k}{(k-1)!}\frac{1}{(2k-1)}.
\en 

Now it is convenient to represent the last fraction as an integral $\frac{1}{2k-1}=\int_0^{1}d\tb \tb^{2k-2},\quad k\ge 1$ and substitute this formula into Eq.(\ref{p_two}) 
\be 
P^{(2)}=\frac{4\sqrt{\pi}}{t}\int_0^N dx \int_0^{1}d\tb \frac{1}{\tb^2} \sum_{k=1}^{\infty}\left(\frac{-\tb^2 b^2t^2}{2x^{2(1-\epsilon)}}\right)^k \frac{k}{(k-1)!}.
\en
Changing $\tb$ by $\beta=\tb bt/\sqrt{2}$ and using the fact that $\sum_{k=1}^{\infty}\left(-y\right)^k \frac{k}{(k-1)!}=-y(1-y)e^{-y}$ we arrive at the following integral representation for the return probability:
\be \label{P2_rep}
P^{(2)}=
\int_0^{\tau}\frac{d\beta}{\beta^2}\int_0^N dx F_2\left(\frac{\beta}{x^{1-\epsilon}}\right),\quad F_2(y)\equiv -2 \sqrt{2\pi}b y^2(1-y^2)e^{-y^2},
\en
where $\tau=bt/\sqrt{2}$. 
%
\section{Scaling exponents: the first order perturbation theory results}\label{sec-exp-first}

The above representation for $P^{(2)}(t)$ (\ref{P2_rep}) is a convenient starting point for calculation of $d_2$ and $\mu$. The exponent $\mu$ can be extracted from the limit $N\to \infty$ of Eq.(\ref{P2_rep})
\be\label{P2_tau}
\Pi_2(\tau)\equiv\lim_{N\to \infty}P^{(2)}=\int_0^{\tau}d\beta\beta^{\frac{2\epsilon-1}{1-\epsilon}}\int_0^\infty dy F_2\left(\frac{1}{y^{1-\epsilon}}\right),\quad y=\beta^{\frac{1}{\epsilon-1}}x.
\en 
Differentiating $\Pi_2(\tau)$ with respect to $\ln \tau$ we obtain
\be\label{J-def}
\dptau=\tau^{\frac{\epsilon}{1-\epsilon}}J,
\quad J\equiv \int_0^\infty dy F_2\left(\frac{1}{y^{1-\epsilon}}\right).
\en
The last step in calculating $\mu$ is to take the limit $\epsilon\to0$ in the expression for $\dptau$. The $\tau$ dependent factor then gives one and what we need to know is just the zeroth order, i.e. $\epsilon$ independent term, in the $\epsilon$-expansion of $J$. The required expansion can be found with the help of the following general formula, which can be proved using the integration by parts:
\be\label{eps-exp}
\int_0^{\infty}d\beta \beta^{\delta-1} f({\beta})=\frac{1}{\delta}f(0)-\int_0^{\infty}d\beta \ln\beta\frac{df}{d\beta}+O(\delta).
\en
To this end we change the integration variable $y$ by $t=1/y^{2(1-\epsilon)}$ and apply the above formula:
\begin{equation}\label{J}
 J=-\frac{\sqrt{2\pi}b}{(1-\epsilon)}\int_0^{\infty} \!\!\! dt 
\frac{(1-t)e^{-t}}{t^{-\frac{1}{2(1-\epsilon)}}}=
 -\frac{\pi b}{\sqrt{2}}\left[1+\epsilon(2+\frac{\gamma}{2}+\ln 2)\right]+O(\epsilon^2),
\end{equation}
where $\gamma$ is the Euler's constant. We keep the first order in $\epsilon$ term, as it is important in the second order perturbation theory. The zeroth order term gives the exponent $\mu$:
\be\label{nu-first}
1-\mu=\frac{\pi b}{\sqrt{2}} +O(b^2).
\en
Now we perform similar calculations for the fractal dimension $d_2$. First we introduce in Eq.(\ref{P2_rep}) the new integration variables  $\tilde{x}$ and $\tilde{\beta}$ defined by the relations $x=\tilde{\beta}^{\frac{1}{1-\epsilon}}N\tilde{x}$, $\beta=N^{1-\epsilon}\tilde{\beta}$ and then take the limit $\tau\to\infty$
\be
\Pi_2(N)\equiv\lim_{\tau\to\infty}P^{(2)}= N^{\epsilon}\int_0^{\infty}d\tilde{\beta}\tilde{\beta}^{\frac{2\epsilon-1}{1-\epsilon}}
\int_0^{\tilde{\beta}^{-\frac{1}{1-\epsilon}}} d\tilde{x}  F_2\left(\frac{1}{\tilde{x}^{1-\epsilon}}\right).
\en
Taking the derivative with respect to $\ln N$ and returning to the previous notation for the integration variables $\tilde{x}\to x$, $\tilde{\beta}\to\beta$ we obtain:
\be
\frac{\partial \Pi_2}{\partial \ln N}= \epsilon N^{\epsilon}\int_0^{\infty}d\beta\beta^{\frac{2\epsilon-1}{1-\epsilon}}
\int_0^{\beta^{-\frac{1}{1-\epsilon}}} dx  F_2\left(\frac{1}{x^{1-\epsilon}}\right).
\en
Then we apply Eq.(\ref{eps-exp}) with  $\delta=\frac{\epsilon}{1-\epsilon}$ and find
\begin{eqnarray}
&&\int_0^{\infty}d\beta\beta^{\frac{2\epsilon-1}{1-\epsilon}}
\int_0^{\beta^{-\frac{1}{1-\epsilon}}} dx  F_2\left(\frac{1}{x^{1-\epsilon}}\right)=\nonumber\\
&&\frac{1-\epsilon}{\epsilon}\left( \int_0^{\infty} dx  F_2\left(\frac{1}{x^{1-\epsilon}}\right)+
\epsilon\int_0^{\infty}d\beta \ln\beta\frac{1}{\beta^2}F_2(\beta)+O(\epsilon^2) \right).
\end{eqnarray} 
The $\epsilon$-expansion of the first integral is given by Eq.(\ref{J}), while the second integral can be calculated explicitly using the definition of $F_2$:
\be
\int_0^{\infty}d\beta \ln\beta\frac{1}{\beta^2}F_2(\beta)=\frac{\pi b}{\sqrt{2}}(1+\gamma/2+\ln 2).
\en
Thus we arrive at the following result for $\frac{\partial \Pi_2}{\partial N}$:
\be
\frac{\partial \Pi_2}{\partial \ln N}=N^{\epsilon}\left[-\frac{\pi b}{\sqrt{2}}+O(\epsilon^2)\right].
\en
It is interesting to note that the first order in $\epsilon$ term is absent in the above expression. The constant term yields the fractal dimension $d_2$ \cite{refME}
\be
d_2=\frac{\pi b}{\sqrt{2}} +O(b^2).
\en
Comparing this result with the corresponding expression for $1-\mu$ Eq.(\ref{nu-first}) we conclude that the Chalker's ansatz is valid in the first order perturbation theory.

Leading contributions  of order of $ \, b \, $ to the scaling exponents $ \, 1 - \mu \, $ and $ \, d_2 \, $ 
in the orthogonal case are calculated in \ref{GOE-2Q}.

\section{Integral representations for $P^{(3)}$} \label{sec-P3}

The second order perturbation result for the return probability can be derived from the corresponding expression for the matrix elements of Green's functions given by Eq.(72) of \pap:
\begin{eqnarray}\label{ret_prob_three}
   P^{(3)} & = & \frac{\pi}{16t^2 N} 
   \sum_{p=1}^N\sum_{ \{m,n \ne p \} }^N \sum_{k_{1,2,3}=0}^{\infty} 
            \left( - 8 b_{pm} t^2 \right)^{k_1}
            \left( - 8 b_{pn} t^2 \right)^{k_2}
            \left( - 8 b_{mn} t^2 \right)^{k_3}  \times \nonumber\\
    & &
             \frac{ \Xi ( k_1, k_2, k_3 ) }{ \Gamma\bigl( 2[k_1+k_2+k_3] - 1 \bigr) } \, 
             ( k_1 + k_2 ) ( k_1 + k_2 - 1 ) ,
\end{eqnarray}
where 
\begin{eqnarray}
   \Xi ( k_1, k_2, k_3 ) & = &\prod_{i=1}^3
              \frac{\Gamma(k_i - 1/2)}{\pi^{1/2} \, k_i! } 
              \times \nonumber\\
   & & 
                 ( 2 k_1 k_2 k_3 - k_1 k_2 - k_1 k_3 - k_2 k_3 ) \, \Gamma(k_1 + k_2 +k_3 -1).
\end{eqnarray}
First we multiply and divide the last expression by $2(k_1+k_2+k_3-1)$ and then use the identity $\Gamma (2z)=\frac{1}{\sqrt{2\pi}} 2^{2z-1/2}\Gamma (z)\Gamma (z+1/2)$ for $z=k_1+k_2+k_3-1/2$, which allows us to cancel $\Gamma(k_1 + k_2 +k_3)$:
\begin{eqnarray}\label{gen}
    P^{(3)} & = & \sum_{k_{1,2,3}=0}^{\infty} F(k_1,k_2,k_3) \times\nonumber\\
&& ( 2 k_1 k_2 k_3 - k_1 k_2 - k_1 k_3 - k_2 k_3 )( k_1 + k_2 ) ( k_1 + k_2 - 1 )            
\end{eqnarray}
with
\begin{eqnarray}\label{G}
   && F ( k_1, k_2, k_3 ) = \frac{\pi}{16t^2N}
   \sum_{p=1}^N\sum_{ \{m,n \ne p \} }^N 
            \left( - 8 b_{pm} t^2 \right)^{k_1}
            \left( - 8 b_{pn} t^2 \right)^{k_2}
            \left( - 8 b_{mn} t^2 \right)^{k_3}\times\nonumber\\
&& \prod_{i=1}^3
              \frac{\Gamma(k_i - 1/2)}{\pi^{1/2} \, k_i! } 
               \frac{\sqrt{2\pi}\;2^{3/2-2(k_1+k_2+k_3)} }{(k_1+k_2+k_3-1)\Gamma(k_1 + k_2 +k_3 -1/2)}.
\end{eqnarray}

All terms in Eq.(\ref{gen}) are symmetric functions of $k_1$, $k_2$ and $k_3$, 
except for the term $(k_1+k_2)(k_1+k_2-1)$. Symmetrizing it we can write $P^{(3)}$ in the following form
\begin{eqnarray}
 P^{(3)} &=& \sum_{k_{1,2,3}=0}^{\infty} F(k_1,k_2,k_3) 
\sum_{\alpha_{1,2,3}}\acoef k_1^{\alpha_1}k_2^{\alpha_2}k_3^{\alpha_3}.
\end{eqnarray}
The coefficients $\acoef$ are invariant under permutations of the indices and hence all non-zero coefficients 
can be obtained from the following six:
\begin{eqnarray}
A_{012}=2/3, \quad A_{013}= -2/3,\quad A_{111}= 2,\nonumber\\
A_{112}= -10/3,\quad A_{113}= 4/3,\quad A_{122}= 4/3.
\end{eqnarray}
 
The next step is to replace the summation over $m$, $n$, and $p$ by the integration similarly to how it was done in the calculation of $P^{(2)}$ . To this end we notice, that 
\begin{eqnarray}\label{sum-int3}
&&\sum_{p\neq m\neq n} f(|p-m|,|p-n|,|m-n|)\approx\nonumber\\
&&6 \int_{0}^N dy\int_0^{y} dx_1 \int_{x_1}^y dx_2 f(|x_1|,|x_2|,|x_2-x_1|)\approx \nonumber\\
&& 6 N \int_0^{N} dq_1 \int_{0}^{N-q_1} dq_2 f(|q_1|,|q_2|,|q_2+q_1|),
\end{eqnarray}
provided that $f(x_1,x_2,x_3)$ is invariant under permutations of the arguments. The last equality is again justified in \ref{app1}. 
In order to be able to sum up over $k_i$ we use the following integral representations:
\begin{eqnarray}\label{inv_gamma}
\frac{1}{\Gamma(k_1+k_2+k_3-1/2)}&=&\frac{1}{2\pi i}\int_{-i\infty+0}^{i\infty+0}ds\: e^s s^{-(k_1+k_2+k_3)+1/2},\nonumber\\
\frac{1}{2(k_1+k_2+k_3-1)}&=&\int_0^1d\tilde{\beta} \tilde{\beta}^{2(k_1+k_2+k_3)-3}.
\end{eqnarray}
The summation over $k_1$, $k_2$ and $k_3$  can be done easily now. All sums over $k_i$ have the form  
$\sum_{k=0}^{\infty} (-y)^k \frac{\Gamma(k - 1/2)}{\sqrt{\pi}\Gamma(k+1)}k^{\alpha}\equiv f_{\alpha}(y),\; \alpha=0,1,2,3$. The explicit expressions for $f_{\alpha}$ are given by
\begin{eqnarray}\label{f}
f_0(y)&=& -2\sqrt{1+y},\quad f_1(y)= -\frac{y}{\sqrt{1+y}},\nonumber\\
f_2(y)&=& -\frac{y(2+y)}{2(1+y)^{3/2}}\quad f_3(y)= -\frac{y(4+2y+y^2)}{4(1+y)^{5/2}}.
\end{eqnarray}
Using this notation and changing $\tilde{\beta}$ by $\beta=\tilde{\beta} bt/\sqrt{2}$ we can write $P^{(3)}$  in a compact form
\be\label{P3_compact}
 P^{(3)}=\int_0^{\tau}\frac{d\beta}{\beta^3}\int_0^{N}dx\int_0^{N-x}dy F_3\left(\frac{\beta}{x^{1-\epsilon}}, \frac{\beta}{y^{1-\epsilon}}\right),
\en
where $F_3$ is defined as 
\begin{eqnarray}\label{F3}
F_3(x,y)&=&\frac{3\pi^{3/2}b^2}{4\pi i}\int_{-i\infty+0}^{i\infty+0}ds e^s \sqrt{s}
	\sum_{\alpha_{1,2,3}}\acoef\times \nonumber\\
 &&f_{\alpha_1}(x^2/s)f_{\alpha_2}(y^2/s)f_{\alpha_3}\left((x^{\epsilon-1}+y^{\epsilon-1})^{2(\epsilon-1)}/s\right).
\end{eqnarray}
Note that Eq.(\ref{P3_compact}) is similar to the integral representation for $ \, P^{(2)} $, Eq.(\ref{P2_rep}). 


\section{Dynamical scaling exponent $\mu$: the second order perturbation theory result}\label{sec-mu}

To calculate the second order result for the dynamical scaling exponents $\mu$  one needs first to find
\be\label{P3_tau}
\Pi_3(\tau)=\lim_{N\to\infty}\left(P^{(3)}-\frac{1}{2}(P^{(2)})^2\right).
\en
For $P^{(2)}$ one can use the representation (\ref{P2_tau}) allowing to integrate over $\beta$ explicitly. To exploit a similar trick for $P^{(3)}$ we scale the integration variables in Eq.(\ref{P3_compact}) $x\to \beta^{\frac{1}{1-\epsilon}}x$, $y\to \beta^{\frac{1}{1-\epsilon}}y$. Then $\tau$-dependence get factorized for both terms in (\ref{P3_tau}) and the derivative with respect to $\ln\tau$ can be now taken explicitly:
\be\label{P3_der_tau}
\frac{\partial \Pi_3}{\partial \ln\tau}=\tau^{\frac{2\epsilon}{1-\epsilon}}\left[
\int_0^{\infty}dx\int_0^{\infty}dy F_3\left(x^{\epsilon-1}, y^{\epsilon-1}\right)-\left(\frac{1-\epsilon}{\epsilon}\right)J^2
\right],
\en 
with $J$ defined in Eq.(\ref{J-def}).

The $\epsilon$-expansion of $J$ is given by Eq.(\ref{J}); its leading term is a constant of order of 
$b$. Therefore $ \, \left(\frac{1-\epsilon}{\epsilon}\right)J^2 \, $ diverges in the limit $ \, \epsilon \to 0 \,$
as $ \, 1/\epsilon $. We show below that the first term in the brackets on the r.h.s. of Eq.(\ref{P3_der_tau}) 
also contains a divergent contribution of order $ \,1/\epsilon \, $, so that two divergent contributions  
{\it cancel}. This cancellation is another manifestation of the existence of the critical dynamical scaling.

Now let us find an $\epsilon$-expansion 
for the double integral $I\equiv \int_0^{\infty}dx\int_0^{\infty}dy F_3\left(x^{\epsilon-1}, y^{\epsilon-1}\right)$. To this end we change variables $x$ by $s^{-\frac{1}{2(1-\epsilon)}}\tilde{x}$, $y$ by $s^{-\frac{1}{2(1-\epsilon)}}\tilde{y}$, 
where $ \, \tilde{x} \, $ and $ \, \tilde{y} \, $ are complex. 
Then the integration  over $s$ in Eq.(\ref{F3}) leads to the appearance of the inverse $\Gamma$-function (\ref{inv_gamma}).  Next we deform the contour of the integration over $\tilde{x}$ and $\tilde{y}$ back to the real axis and get
\begin{eqnarray}\label{G-def}
&&I= \frac{3\pi^{3/2}}{2}b^2 \frac{1}{\Gamma\left(\frac{1}{1-\epsilon}-\frac{1}{2}\right)}
\int_0^{\infty}dx\int_0^{\infty}dy\frac{G(x,y,\epsilon)}{x^{1-\epsilon}y^{1-\epsilon}(x+y)^{1-\epsilon}}\nonumber\\
&&G(x,y,\epsilon)=\sum_{\alpha_{1,2,3}}\acoef g_{\alpha_1}\left(\xx\right)g_{\alpha_2}\left(\yy\right)g_{\alpha_3}\left(\xy\right),
\end{eqnarray}
where the functions $g_{\alpha}$ are related to $f_{\alpha}$ defined in Eq.(\ref{f}) as $g_{\alpha}(y)=f_{\alpha}(y)/\sqrt{y}$. In order to find the $\epsilon$-expansion  of the integral over $x$ and $y$ in the above equation, it is convenient to change the variables by $q=x+y$ and $z=x/(x+y)$. 
In terms of the  new variables the integral, which we denote by $I_0$, takes the form:
\be
I_0=2\int_0^{1/2}dz\int_0^{\infty}\frac{dq}{q^{2-3\epsilon}} \frac{1}{z^{1-\epsilon}(1-z)^{1-\epsilon}} G(qz,q(1-z),\epsilon).
\en
To derive this equation we used the fact that $G$ is a symmetric function, i.e.  $G(x,y)=G(y,x)$. The leading singular term of the $\epsilon$-expansion of $I_0$ originates from $z\to 0$ and can be extracted by integration by parts:
\begin{eqnarray}\label{by_parts}
I_0&=&\frac{2}{\epsilon}\left[\frac{z^{\epsilon}}{z^{1-\epsilon}}
\int_0^{\infty}\frac{dq}{q^{2-3\epsilon}}G(qz,q(1-z),\epsilon)\right|_{z=0}^{z=1/2}-\nonumber\\
&& \left.
\int_0^{1/2}dzz^{\epsilon}\frac{d}{dz}\left\{\frac{1}{(1-z)^{1-\epsilon}}\int_0^{\infty}\frac{dq}{q^{2-3\epsilon}}  G(qz,q(1-z),\epsilon)\right\}\right].
\end{eqnarray}
Thus the $\epsilon$-expansion of $I_0$ has the form $I_0=(2/\epsilon)[A+\epsilon B+O(\epsilon^2)]$. The coefficient $A$ is obtained by setting $\epsilon=0$ in all the terms in the brackets in Eq.(\ref{by_parts}) and it equals to
\be\label{A}
A=\int_0^{\infty}\frac{dq}{q^{2}}G(0,q,0)=\frac{\pi}{6},
\en
where the integral is calculated explicitly using the definition of $G$ (\ref{G-def}). To calculate $B$ we collect the first order terms in $\epsilon$ generated by all $\epsilon$-dependent contributions in Eq.(\ref{by_parts}) and find 
\be
 B=\frac{\pi}{2}-\frac{\pi}{6}\ln 2+R.
\en
The first two constants in the above formula are equal to integrals over $q$ similar to one in Eq.(\ref{A}). The constant $R$ is given by the two-dimensional integral
\be\label{R}
R=\int_0^{1/2}\frac{1}{z}\left[\frac{1}{(1-z)}\int_0^{\infty}\frac{dq}{q^{2}}  G(qz,q(1-z),0)-\frac{\pi}{6}\right]\approx 0.276,
\en
which we were able to compute only numerically. The derived results for $A$ and $B$ along with the $\epsilon$-expansion of the inverse $\Gamma$-function in Eq.(\ref{G-def}) yields
\be
I=\frac{\pi^2b^2}{2\epsilon}\left[1+\epsilon\left(3+\gamma+\ln 2+\frac{6R}{\pi}\right)+O(\epsilon^2)\right].
\en
Substituting this formula as well as the expression for $J$ Eq.(\ref{J}) into Eq.(\ref{P3_der_tau}) we obtain
\be
\frac{\partial \Pi_3}{\partial \ln\tau}=\pi^2b^2\left(\frac{3R}{\pi}-\frac{\ln 2}{2}\right)\approx-0.819b^2.
\en
An alternative integral representation of this answer can be found in Eqs.(22-23) of Ref.\cite{KOYC10}.
We emphasize that the singular $1/\epsilon$ terms  cancel giving a finite result in the limit $ \, \epsilon\to 0 $.
Thus we have demonstrated the existence of the dynamical scaling with the accuracy of the sub-leading
terms of the perturbation theory.
%
%

\section{Fractal dimension $d_2$: the second order perturbation theory result}\label{sec-d2}

To calculate the fractal dimension $d_2$ one needs to deal with 
\be\label{P3_N}
\Pi_3(N)=\lim_{\tau\to\infty}\left(P^{(3)}-\frac{1}{2}(P^{(2)})^2\right).
\en
Changing the integration variables $x=\tilde{\beta}^{\frac{1}{1-\epsilon}}N\tilde{x}$,  $y=\tilde{\beta}^{\frac{1}{1-\epsilon}}N\tilde{y}$  and $\beta=N^{1-\epsilon}\tilde{\beta}$ in the integral representations  (\ref{P2_rep}) and (\ref{P3_compact}) for $P^{(2)}$ and $P^{(3)}$ respectively and taking the derivative with respect to  $\ln N$ we find
\begin{eqnarray}
\frac{\partial \Pi_3}{\partial \ln N}&=& 2\epsilon N^{2\epsilon}\left[\int_0^{\infty}d\beta\beta^{\frac{3\epsilon-1}{1-\epsilon}}\int_0^{\beta^{-\frac{1}{1-\epsilon}}}dx\int_0^{\beta^{-\frac{1}{1-\epsilon}}-x}dy F_3\left(\frac{1}{x^{1-\epsilon}}, \frac{1}{y^{1-\epsilon}}\right)-\right.\nonumber\\
&&\left.\frac{1}{2}\left(\int_0^{\infty}d\beta\beta^{\frac{2\epsilon-1}{1-\epsilon}}
\int_0^{\beta^{-\frac{1}{1-\epsilon}}} dx  F_2\left(\frac{1}{x^{1-\epsilon}}\right)\right)^2\right],
\end{eqnarray}
where we returned to the previous notation for $x$, $y$, and $\beta$. Since we are interested in the limit $\epsilon\to 0$ we can integrate over $\beta$ using the formula (\ref{eps-exp}). In this way we obtain that $\frac{\partial \Pi_3}{\partial \ln N}$ is very similar to the expression (\ref{P3_der_tau}) for $\frac{\partial \Pi_3}{\partial \ln \tau}$:
\be\label{dP_dN}
\lim_{\epsilon\to 0}\frac{\partial \Pi_3}{\partial N}=\lim_{\epsilon\to 0}\frac{\partial \Pi_3}{\partial \tau}+\lim_{\epsilon\to 0}[
2\epsilon K_1(\epsilon)]-K_2,
\en
with
\begin{eqnarray}\label{K1}
 K_1(\epsilon)&=&\int_0^{\infty}d\beta \ln\beta\beta^{\frac{\epsilon-2}{1-\epsilon}}
\int_0^{\beta^{-\frac{1}{1-\epsilon}}}dx F_3\left(\frac{1}{x^{1-\epsilon}}, \frac{1}{(\beta^{-\frac{1}{1-\epsilon}}-x)^{1-\epsilon}}\right),\\
K_2&=&2\int_0^{\infty} dx  F_2\left(\frac{1}{x}\right)\int_0^{\infty}d\beta \ln\beta\frac{1}{\beta^2}F_2(\beta).
\end{eqnarray}
The r.h.s. of Eq.(\ref{dP_dN}) does not contain  divergent contributions of order $ 1/\epsilon \, $, as
divergences cancel out in $ \, \frac{\partial \Pi_3}{\partial \tau} \, $ (see the previous Section) and
$ \lim_{\epsilon\to 0}[2\epsilon K_1(\epsilon)] \, $ is finite (see Eq.(\ref{lim_K_1}) below). This  
fact demonstrates the existence of the spatial scaling $ \, 1/N^{d_2} \, $ with the accuracy of the sub-leading
terms of the perturbation theory. The validity of the Chalker's ansatz implies that $\lim_{\epsilon\to 0}[
2\epsilon K_1(\epsilon)]-K_2=0$ and this is what we show below. 

Let us first calculate $K_2$. Using the explicit expression for $F_2$ (\ref{P2_rep}) one can easily evaluate both integrals
\begin{eqnarray}
&&\int_0^{\infty} dx  F_2\left(\frac{1}{x}\right)=-\frac{\pi b}{\sqrt{2}},\nonumber\\
&&\int_0^{\infty}d\beta \ln\beta\frac{1}{\beta^2}F_2(\beta)=\frac{\pi b}{\sqrt{2}}(1+\gamma/2+\ln 2),
\end{eqnarray}
hence we obtain
\be
K_2=-\pi^2b^2(1+\gamma/2+\ln 2).
\en
For $K_1$ we are interested only in the leading $1/\epsilon$ term of the $\epsilon$-expansion. 
In order to extract it , we first change the variable $\beta$ by $y=\beta^{-\frac{1}{1-\epsilon}}-x$ in Eq.(\ref{K1}) and find that $K_1(\epsilon)=-(1-\epsilon)^2I_1$ with
\be
I_1=\int_0^{\infty}dx\int_0^{\infty}dy\ln (x+y) F_3\left(\frac{1}{x^{1-\epsilon}}, \frac{1}{y^{1-\epsilon}}\right).
\en
The expression for $I_1$ has a structure very similar to the structure of $I$ defined below Eq.(\ref{P3_der_tau}) and hence its $\epsilon$-expansion can be found in exactly the same way. Skipping the details of the calculation we present here only the final result
\be\label{lim_K_1}
\lim_{\epsilon\to 0}[2\epsilon K_1(\epsilon)]=-\pi^2b^2(1+\ln 2+\gamma/2)=K_2.
\en 
Thus we conclude that the contributions of the two last terms in Eq.(\ref{dP_dN}) cancel
 and $\lim_{\epsilon\to 0}\frac{\partial \Pi_3}{\partial N}=\lim_{\epsilon\to 0}\frac{\partial \Pi_3}{\partial \tau}$. This equality not only proves the validity of the Chalker's ansatz in the second order  perturbation theory, but it also provides the expressions for $d_2$ and $\mu$:
\be\label{d2-final}
d_2=1-\mu=\frac{\pi b}{\sqrt{2}}-\pi^2b^2\left(\frac{3R}{\pi}-\frac{\ln  2}{2}\right) +O(b^3),
\en
where $R$ is defined in Eq.(\ref{R}).

\section{Conclusion}

In the above calculations we have demonstrated by
expansion in the parameter $b\ll 1$ that  the power-law scaling
Eq.(\ref{IPR}) and Eq.(\ref{C(omega)}) holds true as soon as $\ln N\gg1$ and
$\ln(E_{0}/\omega)\gg 1$, $ \, E_0 \sim b $, even in the perturbative region where
$b\,\ln N \ll 1$ and $b\,\ln(E_{0}/\omega)\ll 1$. This statement is
verified up to the second order in $b\ll 1$. With the same accuracy
we have shown that the exponents $d_{2}$ and $\mu$ are connected by
the scaling relation Eq.(\ref{Chalk-anz}). Moreover we have found a term $\sim
(\pi\,b)^{2}$ in $d_{2}$ (see Eq.(\ref{d2-final})) which appears to enter with
an anomalously small coefficient $0.083\,(\pi\,b)^{2}$.

However in order to obtain all the above results we used the
analogue of the dimensional regularization, replacing
$(b/(n-m))^{2}$ in Eq.(\ref{Ham}) by $(b/|n-m|)^{2(1-\epsilon)}$ and
setting $\epsilon\rightarrow 0$ at the end of calculation. This
trick is well known in quantum field theory and it is working well for
models whose renormalizability is proven. In other words, it works
well, if it is known that the critical exponents (and the power-laws
themselves) do not depend on the short-range details of the system (e.g. on the
form of the function $\langle |H_{nm}|^{2}\rangle$ in Eq.(\ref{Ham}) which
interpolates between the well defined limits at $n=m$ and $|n-m|\gg
1$). We would like to emphasize here that the renormalizability of
the long-range model studied in this paper is not proven. That is
why it may in principle happen that the results derived in the present work
depend on the regularization scheme. We only know that this is not 
the case in the first order in $b$ where all the integrals can be explicitly calculated
using any other regularization. Whether or not the
universality (independence on the interpolating function) holds in
higher orders in $b$ is an interesting open problem.

\ack
We acknowledge support from the DFG through grant SFB TR-12 and the Nanosystems Initiative 
Munich Cluster of Excellence (OYe), the Engineering and Physical Sciences Research Council, grant 
number EP/G055769/1~(AO). OYe and AO acknowledge hospitality of the Abdus Salam ICTP.

%
%

\appendix 
\section{Integration over the ``center of mass'' coordinate}\label{app1}

Passing from discrete sums to integrals in Eq.(\ref{sum-int2}) and Eq.(\ref{sum-int3}), we replaced the integration over the ``center of mass'' coordinate $y$ by multiplication by $N$. The aim of this Appendix is to justify that step. 

In calculating $P^{(2)}$  we deal with the following integral:
\be\label{I_2N}
I_2(N)=\frac{1}{N}\int_0^N dy\int_0^y dx f(x)\equiv \frac{1}{N}\int_0^N dy F(y),
\en
Our calculations show that the asymptotic behavior of $F(y)$ is given by
\be
F(y)=c\ln y + c_0 +O(1/y),
\en
hence 
\be
I_2(N)=c(\ln N -1)+c_0+\dots=c\ln (N/e)+c_0+\dots\approx F(N/e).
\en
So that replacing $(1/N)\int_0^Ndy$ by one  in Eq.(\ref{I_2N}) is equivalent in the asymptotic limit to replacing $N$ by $N/e$. Now let us show that the same is true in calculation of  $P^{(3)}(t)$. The relevant integral has now the following form:
\begin{eqnarray}
I_3(N)&=&\frac{1}{N}\int_{0}^N dy\int_0^{y} dq_1 \int_{0}^{y-q_1} dq_2 f(|q_1|,|q_2|,|q_2+q_1|)\equiv \frac{1}{N}\int_{0}^N G(y).\nonumber\\
\end{eqnarray}
According to our results the asymptotic behavior of $G$ is given by 
\be
G(y)=d_2\ln^2y+d_1\ln y +d_0+O(1/y),
\en
then substituting this expansion into the definition of $I_3(N)$ we find
\begin{eqnarray}
I_3(N)&=&d_2[\ln^2N-2\ln N+2]+d_1[\ln N-1]+d_0+\dots=\nonumber\\
&=&d_2\ln^2(N/e)+d_1\ln (N/e)+d_0+\dots\approx G(N/e).
\end{eqnarray}
Thus the integration over the ``center of mass'' can be taken into account in calculations of both $P^{(2)}$
and $P^{(3)}$ by scaling the system size. However since we are actually interested in calculation of the scaling exponents the scaling of the system size by a constant factor is irrelevant as it follows from Eq.(\ref{d-log}).


\section{Scaling exponents in the orthogonal symmetry class}\label{GOE-2Q}

The leading contribution to the virial expansion of the return probability in the orthogonal case can be
obtained straightforwardly from the results of Ref.\cite{KYC}:
\begin{equation} \label{OrtRetPr-2}
   P^{(2)}_{orth}(t) =  -\frac{\sqrt{2\pi}}{N}\sum_{n\neq p}^{N}e^{-2b_{pn}t^{2}}
                                                                                   2b_{pn}|t| I_{0}\left(2b_{pn}t^{2}\right) \, .
\end{equation}
Here $ \, I_0 \, $ is the modified Bessel function. We have to calculate the double sum in Eq.(\ref{OrtRetPr-2}) 
with logarithmic accuracy at $ \, b t \gg 1 \, $ and $ \, N \gg 1 $. Therefore the formula for $ \, P^{(2)}_{orth} \, $ 
can be reduced to a single integral (cf. the unitary case):
\begin{equation}\label{RetProb-GOE}
  P^{(2)}_{orth} \simeq -2\sqrt{\pi}\, b\int_l^N\frac{{\rm d}x}{x}e^{-\frac{\tau^{2}}{x^2}}
             \frac{\tau}{x} I_{0}\left(\frac{\tau^{2}}{x^2}\right) , \ \tau \equiv \frac{b t}{\sqrt{2}} \, ;
\end{equation}
where $ \, l \, $ is a finite constant. The dominant contribution to the integral  
 is governed by the region $ \, x < \tau $ where the asymptote of the Bessel 
function can be used:
\[
   I_0(z\gg 1)\approx \frac{e^z}{\sqrt{2\pi z}} \, .
\]
%
Thus we can rewrite Eq.(\ref{RetProb-GOE}) with the logarithmic accuracy as follows
\begin{equation}\label{2Q-Ort-crossover}
    P^{(2)}_{orth} \simeq - \sqrt{2} b \int_{l}^{{\rm min}\left( N, \tau \right)} \frac{{\rm d} x}{x} \, .
\end{equation}
Inserting Eq.(\ref{2Q-Ort-crossover}) into Eq.(\ref{d-log}) we find
\begin{equation}
  1 - \mu = d_2 = \sqrt{2} b + O(b^2) \, ,
\end{equation}
which confirms the Chalker's ansatz up to the leading terms of the perturbation theory.



\section*{References}
\begin{thebibliography}{99}

\bibitem{Wegner} F. Wegner, Z. Phys. B {\bf 36}, 209 (1980).
\bibitem{ChD88} J.T. Chalker, G.J. Daniell, \prl {\bf 61}, 593 (1988).
\bibitem{Chalk} J.T.Chalker, Physica A, {\bf 167}, 253 (1990).
\bibitem{CK07} E.Cuevas, V.E.Kravtsov, Phys.Rev.B  {\bf 76}, 235119  (2007).
\bibitem{HSch94} B. Huckenstein, L. Schweitzer, \prl {\bf 72}, 713 (1994);
     T. Brandes, et al,
     Ann. Phys. (Leipzig) {\bf 5}, 633 (1996);
     K. Pracz, et al,
     J. Phys.: Condens. Matter {\bf 8}, 7147 (1996).
\bibitem{FM-Crit}
Y.V. Fyodorov, A.D. Mirlin, Phys. Rev. B {\bf 55}, R16001 (1997).
\bibitem{SupCond}
M.V. Feigel'man, et al, \prl {\bf 98}, 027001 (2007); 
M.V.Feigelman, et al, Annals of Physics {\bf 325},
1368 (2010).
\bibitem{Kondo}
S. Kettemann, E. R. Mucciolo, I. Varga, \prl {\bf 103}, 126401 (2009).
\bibitem{FOR09} Y. V. Fyodorov, A. Ossipov, A. Rodriguez, J. Stat. Mech. (2009) L12001.
\bibitem{ROF11} I. Rushkin, A. Ossipov, Y. V. Fyodorov, J. Stat. Mech. (2011) L03001.
\bibitem{MFD96} A.D. Mirlin, et al,
   \pre {\bf 54}, 3221 (1996).
\bibitem{KOYC10} V. E. Kravtsov, A. Ossipov, O. M. Yevtushenko, E.  Cuevas,  \PR B {\bf 82}, 161102(R) (2010).
\bibitem{YK03} O. Yevtushenko, V.E. Kravtsov J.Phys.A: Math.Gen. {\bf 36},
   8265 (2003); Phys. Rev. E {\bf 69}, 026104 (2004).
   V.E. Kravtsov, O. Yevtushenko, E. Cuevas, J.Phys.A: Math.Gen. {\bf 39}, 2021(2006).
\bibitem{Levitov} L.S. Levitov, \prl {\bf 64}, 547 (1990).
\bibitem{YO} O. Yevtushenko and A. Ossipov, J. Phys. A {\bf 40}, 4691 (2007).
\bibitem{KYC} S. Kronm{\"u}ller, O.M. Yevtushenko, E. Cuevas, J. Phys. A: Math. Theor. {\bf 43}, 075001 (2010).
\bibitem{refME} The first order result for $d_2$ was derived in \cite{ME00}.
\bibitem{ME00} A.D. Mirlin, F. Evers, \prb {\bf 62}, 7920 (2000).
\end{thebibliography}
\end{document}